\documentclass[final,authoryear,5p,times,twocolumn]{elsarticle}

\usepackage{graphicx}

\usepackage{amssymb}

\journal{Icarus}

\begin{document}

\begin{frontmatter}



\title{Dust Ejection from Planetary Bodies by
Temperature Gradients: Laboratory Experiments}
\author[due]{Thorben Kelling}
\author[due]{Gerhard Wurm}
\author[miro]{Miroslav Kocifaj}
\author[jozef]{Jozef Kla\v{c}ka}
\author[dennis]{Dennis Reiss} 
\address[due]{Faculty of Physics, Universit\"{a}t Duisburg-Essen, Lotharstrasse 1, 47057 Duisburg, Germany}
\address[miro]{Astronomical Institute, Slovak Academy of Sciences, Dubravska 9, 845 04 Bratislava, Slovak Republic}
\address[jozef]{Faculty of Mathematics, Physics, and Informatics, Comenius University, Mlynsk\'{a} dolina, 842~48 Bratislava, Slovak Republic}
\address[dennis]{Institut f\"{u}r Planetologie, Westf\"alische Wilhelms-Universit\"at M\"{u}nster, Wilhelm-Klemm-Strasse 10, 48149 M\"{u}nster, Germany}

\begin{abstract}
Laboratory experiments show that dusty bodies in a gaseous environment eject dust particles if they are illuminated. 
We find that even more intense dust eruptions occur when the light source is turned off.
We attribute this to a compression of gas by thermal creep in response to the changing temperature gradients in the top dust layers.
The effect is studied at a light flux of $13$ kW/m$^2$ and 1 mbar ambient pressure. The effect is applicable to
protoplanetary disks and Mars. In the inner part of protoplanetary disks, planetesimals can be eroded especially at the terminator
of a rotating body. This leads to the production of dust which can then be transported towards the disk edges or the outer disk regions. The generated dust might 
constitute a significant fraction of the warm dust observed in extrasolar protoplanetary disks. We estimate erosion rates of about 1 kg s$^{-1}$ for 100 m parent bodies. The dust might also contribute 
to subsequent planetary growth in different locations or on existing protoplanets which are large enough not to be susceptible to particle loss by light induced ejection. Due to the ejections, planetesimals and smaller bodies will be accelerated or decelerated and drift outward or inward, respectively. The effect might also explain the entrainment of dust in dust devils on Mars, especially at 
high altitudes where gas drag alone might not be sufficient.
\end{abstract}

\begin{keyword}
planetary systems: formation \sep planetary systems: protoplanetary disks \sep planets and satellites: general \sep mars \sep mars, surface
\end{keyword}

\end{frontmatter}


\section{Introduction}

Dust aggregates composed of $\mu$m sized grains play a major role in several astrophysical processes. The early stages of planet formation, e.g., are based on colliding dust particles and the accretion of dust \citep{dominik2007,blum2008}. While dusty bodies in principle might grow to large sizes through collisions \citep{teiser2009}, meter-sized bodies in protoplanetary disks drift inward rapidly at about 1 AU in 100 years under typical nebular conditions \citep{weidenschilling1977}. Eventually, they might evaporate and be accreted by their host star. If accreted by the star, solids are lost very fast compared to the lifetime of $10^{7}$ years of a protoplanetary disk. 

The dust ejection mechanisms presented in this article work at the magnitude of pressure and light flux found in inner regions of protoplanetary disks ($\ll 1$ AU, mbar pressure and light fluxes $>10$ kW/m$^2$, \citep{wood2000}) and provide methods to disassemble but reintroduce the material of inward drifting larger bodies. The mass of these bodies is therefore not lost through accretion but is recycled and can serve as reservoir of small dust material within the disk.

Smaller ($< 100$ $\mu$m) dust particles dominate the visible and near infrared emission spectra of the warm inner part ($< 10$ AU) of protoplanetary disks and exist for millions of years \citep{olofsson2010,mamajek2004}. In common planet formation models the dust particles readily collide, form aggregates and the smaller dust particles are depleted. The presence of small dust particles observed over the whole lifetime of protoplanetary disks therefore requires mechanisms to replenish the dust by recycling at least parts of the larger bodies. Collisional disruption is one source but by far not the only one. \citet{paraskov2006} argue that planetesimals on slightly eccentric orbits can be eroded in a few orbits due to aeolian erosion by strong head winds. In cases where the gas density is not high enough for aeolian erosion but where the disk is optically thin the mechanism described here can also efficiently erode dusty bodies. This is important for transitional disks which have observable inner transparent gaps sometimes partially filled with gas \citep{sicilia2006,calvet2002}.

Another research field where the physics of dust is important is Mars where different kinds of dust activities are frequently observed. Besides (global) dust storms, active dust devils are common on Mars \citep{balme2006,thomas1985}. The usual explanation is the pick-up of dust by gas drag (winds and vortices) in analogy to dust devils on Earth. However, wind velocities seem to be too low (in general $v<30$ m/s) to explain dust lifting from the martian surface by gas drag only \citep{stanzel2008,greeley1980,ryan1978}. More critically, dust devils are observed even at elevations $>10$ km such as inactive volcanoes where the corresponding atmospheric pressure is as low as $p\sim2$ mbar \citep{reiss2009,cushing2005}.  The low gas pressure requires even higher wind pick-up velocities if gas drag is responsible for particle entrainment. The dust lifting process presented in this article might support dust devil activity on Mars at mbar pressure. Dust devils - once initiated - can be self-sustaining by this
mechanism.

\begin{figure}
\centering
\includegraphics[width=.45\textwidth]{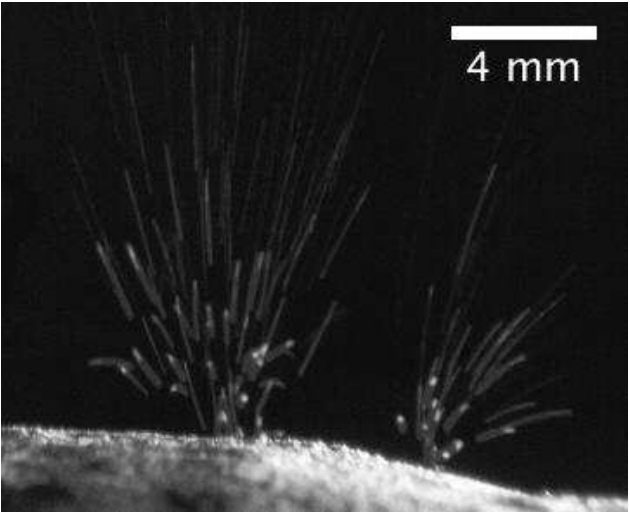}
\caption{Particle eruptions are visible in the dimming light after illumination is switched off (contrast enhanced). The light source illuminated the dust bed surface (grey basalt, bottom) from the top for approximately one minute before. Particle ejections, distinct in space and time, occur for several seconds due to the overpressures induced by the Knudsen compressor effect.\label{fig:basaltblast}}
\end{figure}

The erosion of dusty surfaces by illumination was suggested by \citet{wurm2006}, applied to protoplanetary disks by 
\citet{wurm2007} and to Mars by \citet{wurm2008}. However, those particle ejections were based on photophoretic forces directly acting on the topmost particles. Here, we find in addition a different and potentially more powerful mechanism, the creation of a subsurface overpressure based on thermal creep. Thermal creep is the non-equilibrium gas flow in temperature gradients at (particle) surfaces. It is the counterpart to photophoretic and thermophoretic motion of free solid particles. In a dust bed gas flow due to thermal creep occurs through the pore space. If the pores are small enough that gas cannot flow back efficiently this leads to a build-up of pressure that eventually results in dust eruptions (see e.g. Fig.\ref{fig:basaltblast}).

\begin{figure}
\centering
\includegraphics[width=.45\textwidth]{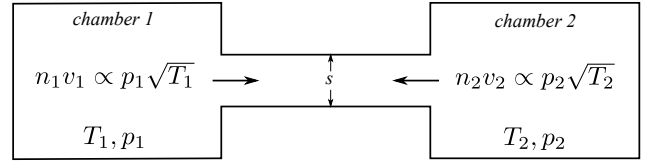}
\caption{Principle of the Knudsen compressor. If two chambers 1 and 2 at different temperatures $T_1, T_2>T_1$ are joined by a connection with diameter $s\ll \lambda$, an overpressure on the warmer side is established with $p_2/p_1=\sqrt{T_2/T_1}$.\label{fig:knudsenprinciple}}
\end{figure}

Compression by thermal creep -- also known as thermal transpiration -- was described and experimentally demonstrated by \citet{knudsen1909}, who built a multi-stage vacuum pump working at low pressure with no moving parts. In his experiment, Knudsen connected two vacuum chambers at temperatures $T_1$ and $T_2>T_1$ with a channel of diameter $s$ that is small compared to the mean free path $\lambda$ of the gas molecules or haslarge Knudsen numbers $Kn\gg 1$ where we define $Kn=\lambda/s$ (see also Fig.\ref{fig:knudsenprinciple}).

He found that in equilibrium the pressure in the two chambers is given by
\begin{equation}
\frac{p_2}{p_1}=\sqrt{\frac{T_2}{T_1}}
\end{equation}
Hence, the warmer chamber is at a higher pressure than the colder chamber not due to thermal expansion of the gas but because of thermal creep from the cold reservoir. \citet{muntz2002} showed, that in the transition regime ($Kn\sim 1$) the overpressure $\Delta p =\mid p_2 -p_1\mid$ is
\begin{equation}
\Delta p = \frac{Q_T}{Q_P}\frac{p_{avg}}{T_{avg}}\Delta T\label{eq:overpressure}
\end{equation}
Here, $Q_T$ and $Q_P$ are thermal creep and Poiseuille flow coefficients of the capillary connection depending on the Knudsen number for the average pressure, $\Delta T = \mid T_2-T_1 \mid$ is the temperature difference over the connection, $T_{avg}=(T_1+T_2)/2$ and $p_{avg}$ are the average temperature and pressure, respectively. \citet{kelling2009} demonstrated that this Knudsen compressor effect is capable of letting dust aggregates levitate over a hot surface. The pores of the dust aggregate act as a collection of microchannels (capillaries) between the bottom and the top surface of the aggregate.

Photophoresis as a dust ejection mechanism for the single top-most surface particles from a dusty body was demonstrated by \citet{wurm2006}. Photophoretic forces or radiometer forces have been known since the late 19th century \citep{reynolds1876,maxwell1879}, and the concept was further developed by \citet{einstein1924}. Photophoresis acts on small particles with a temperature gradient over their surface in a low pressure gaseous environment. In general, photophoresis acts in the direction from warm to cold. \citet{brueche1931} first found a simple formula for the photophoretic force for all pressures which was then explained in detail for spherical particles by \citet{rohatschek1995}. According to \citet{kantorovich1999}, the photophoretic force for $Kn<1$ can be approximated by
\begin{equation}
F_{ph}=\frac{3\pi \eta^2 a}{2 T_{avg}\rho_g}\frac{dT}{dz}\label{eq:photopho},
\end{equation}
with $\eta=1.8\times10^{-5}$ kg/(ms) as the gas viscosity, $a$ as the particle radius, $T_{avg}$ as the average gas temperature, $\rho_g\simeq 0.5\times 10^{-3}$ kg/m$^3$ as the gas density (at 1 mbar and $T=700$ K) and $dT/dz$ as the temperature gradient over the particle's surface.

\section{Experiments}

\begin{figure}
\centering
\includegraphics[width=.45\textwidth]{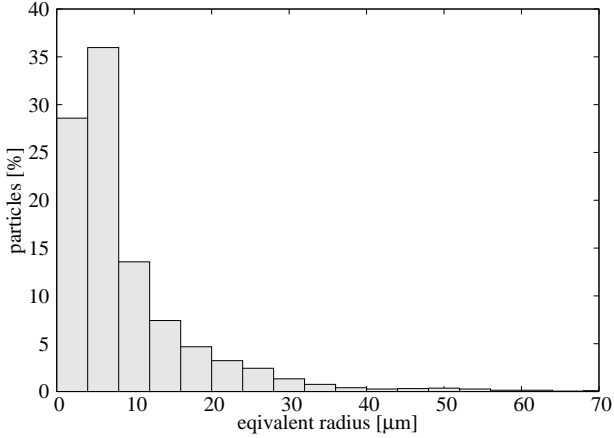}
\caption{Size distribution of the basalt powder (equivalent radii)  used in the experiments.\label{fig:basalt_original}}
\end{figure}

\begin{figure}
\centering
\includegraphics[width=.45\textwidth]{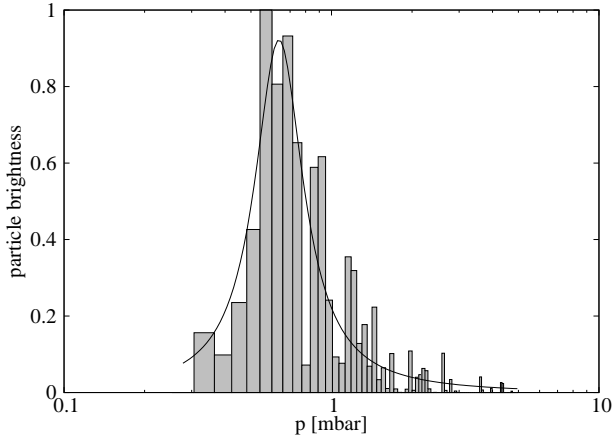}
\caption{Particle release from a basalt sample with illumination while the pressure is varied (constant air inflow of $\sim 0.06$ mbar/s, $I\simeq 13$ kW/m$^2$). The boxes represent the brightness of the ejected particles per time and the solid line is a photophoretic fit through the data points according to Eq.(\ref{eq:roha}). The particle brightness is assumed to be proportional to the amount of ejected particles. From the fit the pressure at maximum particle release is about 0.6 mbar. \label{fig:basalt_pressure}}
\end{figure}

Basalt powder (single grain components $< 100$ $\mu$m) as a dust sample is placed within a vacuum chamber (Fig.\ref{fig:basaltblast}, Fig.\ref{fig:basalt_original}).
The pressure is adjusted to $0.1-10$ mbar while most experiments were carried out around the pressure of $\sim 1$ mbar where the particle ejection rate is at maximum (see below).  A light source (halogen lamp) with a broad visible spectrum is placed outside the vacuum chamber. The lamp is focused onto a surface area of $\sim 1$ cm$^2$ with an intensity of $I\simeq 13$ kW/m$^2$. If the intensity exceeds a certain limit (depending on the dust sample and the pressure, typically $\sim 10$ kW/m$^2$ at mbar pressure), continuous particle ejections on the order of a few particles per second are observable - this is also reported by \citet{wurm2006} but with a significantly higher light flux on a smaller surface spot. The pressure dependence of the continuous particle ejections is depicted in Fig.\ref{fig:basalt_pressure}.

The particle ejections peak around 0.6 mbar and decrease rapidly to lower and higher pressures.

\begin{figure}
\centering
\includegraphics[width=.45\textwidth]{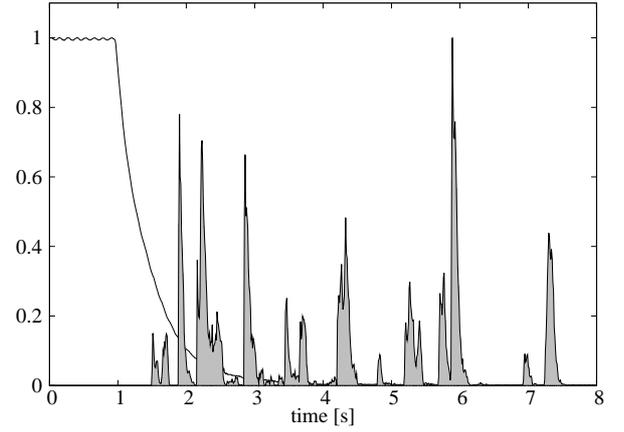}
\caption{Eruptions if the light source is switched off. The solid decreasing line is the normalized light flux incident on the dust bed's surface and the peaks are the normalized particle brightnesses.  About 100 particles per second are released, but only for a short time scale (in general $< 10$ s).\label{fig:knudsenejections}}.
\end{figure}

After an initial illumination phase (approx. one min), the light source is switched off. More numerous particle releases (up to 100/s) from the surface of the dust bed into the surroundings are observed right after the intensity changed. After some seconds these eruptions vanish. In general, the eruptions after the light switch off are distinct events in space and time within the spot that was illuminated before (see Fig.\ref{fig:basaltblast} for the case that two eruptions occur at the same time). The normalized light curve (brightness of background) and particles releases (brightness of particles) are depicted in Fig.\ref{fig:knudsenejections}.

The released aggregates are in general smaller than $100$ $\mu$m in diameter.

\section{Model}

\begin{figure}
\centering
\includegraphics[width=.45\textwidth]{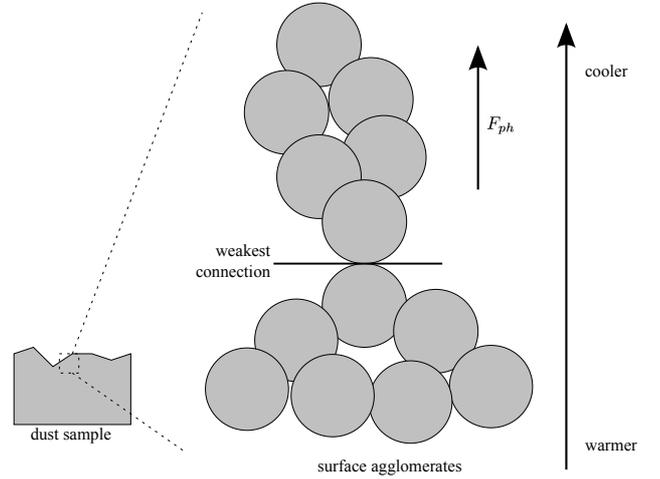}
\caption{Photophoretic ejections (principle). The continuous ejections of single particle aggregates are caused by photophoresis. If the induced photophoretic force $F_{ph}$ overcomes gravity $F_G$ and the cohesion forces at the weakest connection, a surface aggregate is released into the surroundings.\label{fig:photopho}}
\end{figure}

\begin{figure}
\centering
\includegraphics[width=.45\textwidth]{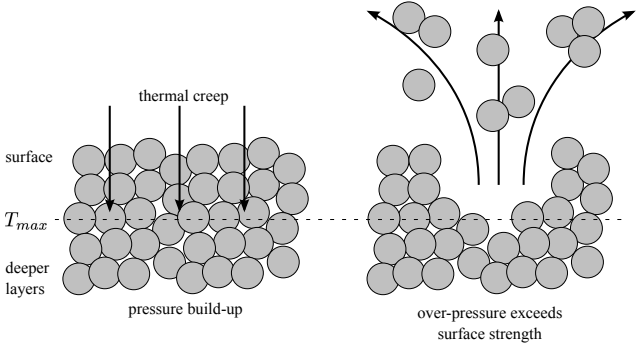}
\caption{If the light source is switched off, $T_{max}$ moves towards deeper dust layers. If the temperature gradient $dT/dz$ covers enough dust layers, gas is efficiently draged downwards by thermal creep towards $T_{max}$ and builds up an overpressure $\Delta p = f\cdot \Delta T$. As soon as the pressure below the surface exceeds the local tensile strength an eruption occurs. For a few seconds this is about 100 times more efficient in particle ejection rate than photophoretic eruptions.\label{fig:knudsensketch}}
\end{figure}

We attribute the continuous particle release while the illumination is on to photophoretic forces acting on the top most dust particles (see Fig.\ref{fig:photopho}, \citealt{wurm2006}) and the more numerous particle release after switching the light off to a surface break-up by a Knudsen compressor like overpressure (Fig.\ref{fig:knudsensketch}).

\begin{figure}
\centering
\includegraphics[width=.45\textwidth]{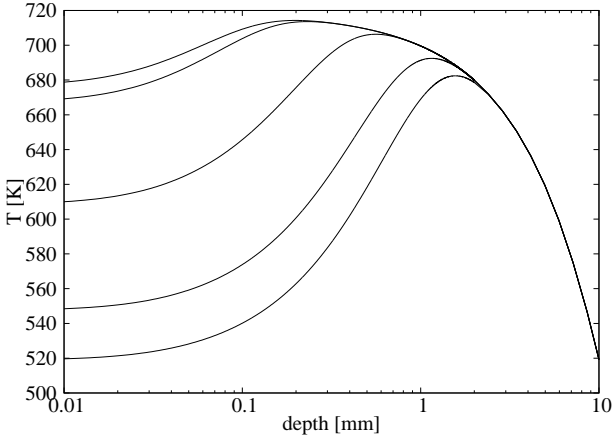}
\caption{Calculations of the temperature within a dust bed (details can be found in \citealt{miro2010b}). The solid lines show the temperature within a dust bed after a $I=10$ kW/m$^2$ light source illuminated the dust bed and is off for 0.05s, 0.1s, 1s, 5s and 10s (top to bottom). With time, the maximum temperature moves deeper into the dust bed. While the temperature gradient towards the surface gets smaller the absolute temperature differerence gets larger for a limited time.\label{fig:greenhouse}}
\end{figure}

The visible light incident on the dust bed's surface heats the upper dust layers due to absorption. While the surface of the dust bed cools by thermal radiation, the deeper layers transport the heat mainly through conduction -- consequently, the maximum temperature is below the surface of the dust bed \citep{miro2010}. This is analogous to the well known greenhouse effect. Two temperature gradients with different sign exist then. The temperature drops from the maximum temperature towards the cooler and deeper inside of the dust bed and it drops toward the cooler surface. We carried out dedicated heat and radiative transfer calculations to determine the temperature gradients, which was the subject of previous publications \citep{miro2010,miro2010b}. At the intensities $I\simeq 13$ kW/m$^2$ used in our experiments, the simulations show, that typically positions of the temperature maximum are at a few $100$ $\mu$m depth within the dust bed (Fig.\ref{fig:greenhouse}, \citealt{miro2010,miro2010b}).

Due to the thermal cooling and the reversed temperature gradient in the upper dust layers during the continuous illumination, photophoresis (Eq.\ref{eq:photopho}) acts on the particles in the direction from warm to cold and hence away from the surface. If the photophoretic force overcomes gravity and cohesion forces, particles are ejected from the dust bed.

According to \citet{miro2010b}, temperature gradients of $10^5$ K/m occur at $10$ kW/m$^2$ illumination. Using Eq.(\ref{eq:photopho}) with $a\simeq 50$ $\mu$m, $\rho_g=0.5\times 10^{-3}$ kg/m$^3$ (at 1 mbar and $T=700$ K) and $T_{avg} = 700$ K (\citep{miro2010b}) one gets a ratio of
\begin{equation}
\frac{F_{ph}}{F_G}\simeq 15
\end{equation}
where $F_G= (4/3) \pi a^3 \rho$ with $\rho = 2900$ kg/m$^3$. This is a sufficient condition to eject particles. If a filling factor of 0.3 of the ejected particles is assumed, the condition would increase to $F_{ph}/F_G\simeq 50$.  Because photophoresis is pressure dependent, the number of particles ejected should correlate to this pressure dependence which is given by \citep{rohatschek1995}
\begin{equation}
F_{ph}=\frac{(2+\delta)F_{max}}{\left(\frac{p}{p_{max}}+\delta+\frac{p_{max}}{p}\right)}\label{eq:roha}
\end{equation}
with $p_{max}$ as the pressure where the maximal photophoretic force $F_{max}$ appears. Both coefficients are extensively discussed in \citet{rohatschek1995} and depend on the gas and particle properties; $\delta$ is a factor which takes the value $\delta\simeq -1.9$ for our normalized data. Fig.\ref{fig:basalt_pressure} depicts the pressure dependence of the number of ejected particles over a pressure range of 0.3--5 mbar while the light source is on. The solid line represents a three parameter fit ($p_{max}$, $F_{max}$, $\delta$) of Eq.(\ref{eq:roha}). The curve fits the normalized data well with most particles being ejected around 0.6 mbar. 

The thermophoretic force $F_{th}$ for $a=50$ $\mu$m particles and $\lambda<a$ is \citep{zheng2002}
\begin{equation}
F_{th}=\frac{f_{th}a^2\kappa_g}{\sqrt{2k_B T_{avg}/m_g}}\frac{dT}{dz}\label{eq:zheng}
\end{equation}
where $f_{th}\simeq 0.02$ is the dimensionless thermophoretic force, $\kappa_g=0.01$ W/mK is the thermal conductivity of the gas, $k_B=1.38\times 10^{-23}$ J/K is the Boltzmann constant and $m_g=4.8\times 10^{-26}$ kg is the molecular mass of air. This yields a ratio of
\begin{equation}
\frac{F_{th}}{F_G}\simeq 10^{-2}.
\end{equation}
These calculations are estimates but thermophoresis is orders of magnitudes too weak to lift particles from the surface while photophoresis is strong enough and the data fit the pressure dependence of photophoresis very well. 

If the light source is switched off after illumination, more frequent and numerous particle releases occur for some seconds until no more particles are ejected. At mbar pressure the mean free path $\lambda$ of the gas molecules ($0.1-10$ mbar corresponds to $\lambda \simeq 700-7$ $\mu$m) is comparable to the mean pore size of the dust bed ($Kn\sim 1$). Our earlier calculations \citep{miro2010,miro2010b} show that the temperature maximum moves deeper into the dust bed
if the light is turned off (Fig.\ref{fig:greenhouse}). If the temperature increases with depth over enough dust layers, the dust bed with its pores acts as a collection of mircochannels. According to Eq.(\ref{eq:overpressure}) an overpressure is established by sucking gas from above the dust bed towards the maximum temperature. The overpressure causes -- if strong enough -- particle ejections (Fig.\ref{fig:knudsensketch}). \citet{kelling2009} demonstrated that such an overpressure induced by thermal creep can levitate dust aggregates over a hot surface.

\begin{figure}
\centering
\includegraphics[width=.45\textwidth]{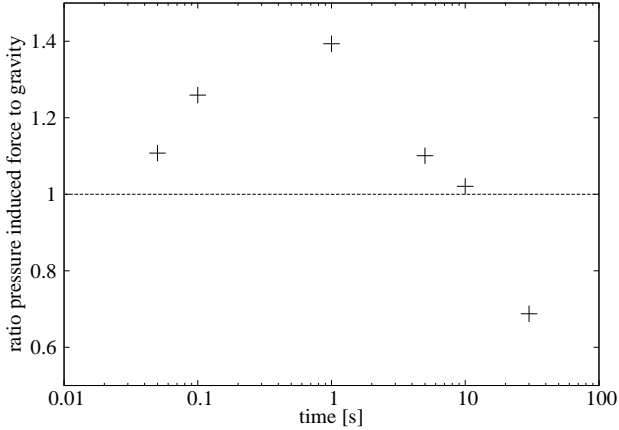}
\caption{Based on previous calculations (\citealt{miro2010b} and Fig.\ref{fig:greenhouse}) the ratio of the overpressure induced force to earth gravity (see text) at a depth of $d=100$ $\mu$m within the dust sample after the illuminatin is switched off ($t=0$) is depicted. Shortly after the switch off the overpressure increases due to rapid surface cooling and hence a greater temperature difference between the considered depth and the surface. It gets strong enough to lift a 100 $\mu$m thick layer (ratio $>1$) for several seconds. The overpressure then decreases with time due to the general cooling and particle release should stop after 
about 10 s. This is in agreement with the experiments. \label{fig:pkn}}
\end{figure}

As the maximum temperature moves deeper into the dust bed, the temperature gradient gets smaller while the absolute temperature difference between the maximum and the surface increases for a certain time scale (from approx. 20 K at continuous illumination to 100 K 1s after the light is switched off).  To raise a 100 $\mu$m layer of basalt against gravity $\Delta p\sim 1$ Pa is sufficient with the general condition
\begin{equation}
\frac{\Delta p A}{F_G}=\frac{\Delta p}{d\rho f g}>1
\end{equation}
with $\Delta p$ as the temperature difference induced overpressure (which is obtained according to Eq.(\ref{eq:overpressure})), $A$ is an area (e.g. the corresponding illuminated area under which the overpressure is present), $F_G=Ad\rho f g$ is the gravitational force, $d=100$ $\mu$m is the thickness of the dust layer, $\rho=2900$ kg/m$^3$ is the density of the dust, $f=0.3$ is the filling factor and $g=9.81$ m/s$^2$ is the gravitational acceleration. After some seconds, the temperature differences are too low to induce a sufficient overpressure -- hence, all dust ejection activity comes to rest on the order of some seconds after the illumination is switched off (in general $< 10$ s). Fig.\ref{fig:pkn} shows the ratio of the overpressure induced force and gravity at a depth of 100 $\mu$m within the dust sample after the light is switched off. 

Very shortly after the switch off of the light source the strength of the overpressure is sufficient to overcome gravity and to eject particles. Taking into account that the overpressure needs some time to build up, the ejections should start with some delay. After some seconds the overpressure is too weak to eject particles fom the dust bed (ratio $<1$,  Fig.\ref{fig:pkn}).

Roughly 100 times more particles per second are released during the eruptions after the light is switched off compared to the continuous photophoretic ejections -- but for a limited time.

The simulations (\citealt{miro2010,miro2010b} and Fig.\ref{fig:greenhouse}) are based on a homogenous dust slab of one particle size (5 $\mu$m). Also the photophoretic and thermophoretic forces in the calculations are only determined for spherical particles of one size. In the experiments the dust bed is composed of particles with a wider size distribution (Fig.\ref{fig:basalt_original}) and of varying shapes which may increase or decrease the absolute forces. The light source covers only a small region on the dust bed's surface. While the extend is large ($\sim 1$ cm) compared to the relevant depths and individual particle sizes ($< 1$ mm) we cannot exclude edge effects, i.e. gas leakage to the sides. While this might rather decrease any overpressure in favour for the effect for more extended illuminated surfaces, the influence of discrete particle size and illumination needs future work beyond the scope of this paper.

\section{Applications}
\begin{figure}
\centering
\includegraphics[width=.45\textwidth]{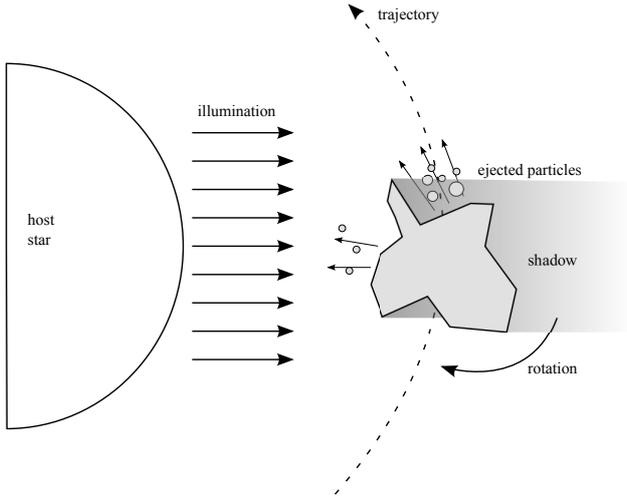}
\caption{Sketch of ejections. On the illuminated side the parent body continuously ejects particles. Uneveness on a rotating body can cause a rapid change from light to shadow. At the terminator the more massive eruptions will occur. Depending on the rotation and the uneveness the ejected particles might change the trajectory of the parent body.\label{fig:planet_sketch}}
\end{figure}

The evolution of planetary bodies in the inner part of protoplanetary disks might be a key process to understand dust observations, thermal processing in protoplanetary disks or local planet formation.
As an example, meter-size bodies in typical disk scenarios drift rapidly inward. They would be accreted by their host star within some 100 years if they are not processed by some means before. As collisions will continuously produce objects in that critical size range there is always a net inward mass flow of solids. This is part of the reason why this size range is often named the meter-size barrier \citep{brauer2008}. The dust ejections presented in this article might be one mechanism to prevent the mass loss of solids in a protoplanetary disk. We propose that larger bodies are (partially) eroded and the ejected matter is recycled by re-injecting the material into the disk. In a recent paper \citet{wurm2007} considered the destruction of illuminated dusty bodies. Here, we showed that the transition to darkness is an even more effective mechanism of destruction. In the optical thin part of the disk close to a star a change from illumination to darkness always occurs by e.g. uneveness (craters, ridges) on rotating bodies close to the terminator (Fig.\ref{fig:planet_sketch}).

As the bodies drift inwards, the light flux of the star easily exceeds 10 kW/m$^2$. Pressures in the mbar regime might be present \citep{wood2000}. Similar to the presented experiments, light-shadow induced blasts of dust particles will occur in protoplanetary disks. It has to be noted that gravity and cohesion has to be matched in the laboratory experiments to eject matter. For meter-size bodies or even larger planetesimals self-gravity is not significant and ejections should require less stringent conditions, i.e. a wider pressure range or lower light flux. 
In addition, the high intensities close to the star also might induce continuous photophoretic ejections \citep{wurm2007}.

Eroded material is re-injected into the protoplanetary disk and is transported outwards and upwards where it can be observed, e.g. as warm dust on the surface of the disk. Transport might be provided by photophoresis \citep{wurm2009} or turbulence \citep{takeuchi2003,ciesla2007}. 
The dust might take part in further planet formation. Already existing planetary bodies not susceptible to particle loss any more due to sufficient self-gravity might accrete the dust. This way, the growth of planetary bodies would benefit from the destruction of smaller bodies which would otherwise not be within their gravitational reach or harder to be accreted than the dust.
 
Our experiments show that 100 particles per second with a radius of 50 $\mu$m are ejected from a 1 cm$^2$ spot if the light is dimmed on a timescale of seconds. Assuming a mass density of 2900 kg m$^{-3}$, a filling
factor of 0.3 this is approx. $10^{-3}$ kg s$^{-1}$ m$^{-2}$ but only for about 10s, which gives a total mass ejected per area of
$10^{-2}$ kg m$^{-2}$. We assume that such mass loss can occur close to the terminator of 
a (rapidly) rotating body where ridges or craters cast pronounced shadows (Fig.\ref{fig:planet_sketch}). This way,
in a simplified model, the total surface of the body is subject to eruptions once upon every rotation. The total mass
loss will depend on the size of the object, the rotation rate and -- in detail -- on the surface morphology.
E.g. for an $r=100$ m radius body, rotating once every 10 hours the total mass loss is $10^{-2}$ kg s$^{-1}$. 

The mass loss rate per area of the photophoretic ejections by direct illumination is about one particle per second and
cm$^2$ which is $10^{-5}$ kg s$^{-1}$ m$^{-2}$. For an illuminated body in a protoplanetary disk in a simple model it acts on one hemisphere continuously. Therefore the mass loss for the 100 m body considered above is 1 kg s$^{-1}$. 
This is much larger than the mass loss due to compressed gas. However, it has to be noted that 
the ejection mechanisms are different and it is not clear that 
both mechanisms work equally well close to the threshold in light flux where eruptions might start.
Another important difference is the influence on the bodies trajectory. In principle the parent body of the released particles has to balance the momentum loss due to the ejected particles. The photophoretic eruptions along the 
direction of illumination (from the star) imply a radial force. This does not directly change the orbit but reduces the 
orbital velocity of the body. Subsequently gas drag of the disk might decelerate the body. However, the
ejections at the terminator directly imply a force along the orbit.  Depending on the direction of rotation the body will loose mass at the front or back side and be decelerated or accelerated, respectively (Fig.\ref{fig:planet_sketch}). This will cause the object to drift inwards or outwards. For outward moving bodies this might imply that a concentration distance occurs where the illumination is just at the threshold for particle eruptions. While the speed by which this distance is reached should depend on the rotation rate, the actual distance should be insensitive to the rotation rate but only depend on the limit for particle eruptions.

On Mars, dust entrainment is an unsolved problem, especially as dust devils are observed at high elevations of 10 km and more ($p\sim 2$ mbar, \citealt{reiss2009}). Gas drag seems to be too weak to explain dust lifting then \citep{greeley1980}. The presented ejection mechanisms in contrast work best at mbar pressure. As indicated by \citep{wurm2008}, photophoretic ejections are possible on Mars. Dust devils on Mars are optically thick. Shadows of dust devils are observed from space \citep{cushing2005}. Hence, if dust devils translate over the surface they reduce the infalling flux on the ground. The surficial dust layer on Mars seems to consist of weakly-bound dust aggregates \citep{sullivan2008}. Dust particle eruptions might occur, feeding the dust devil with particles. Dust devils on Mars might therefore be self-sustaining, as long as the needed eddy -- unnoticed without dust -- exists and if initiated by any means. The light flux on Mars reaches values of $I\sim 700$ W/m$^2$. Temperature differences for the Knudsen induced eruptions are therefore less on Mars then for planetesimals close to a star in protoplanetary disks. It is not clear how much mass can be ejected under martian conditions by this mechanism and we currently cannot unambiguously claim that the mechanism described is responsible for dust lifting within dust devils. At minimum it might support other mechanisms. Future work has to show if relevant mass rates can be produced at martian light intensities. Nevertheless, there is little doubt that the gas flow associated with the Knudsen compression is working efficiently within the regolith of Mars and might contribute to other topics like (water) vapor transport.

\section{Conclusion}
We showed, that particles are released from illuminated dust beds at mbar pressures continuously (Fig.\ref{fig:basalt_pressure}) and that even more particles are ejected from the surface after the illumination is switched off (Fig.\ref{fig:knudsenejections} and Fig.\ref{fig:basaltblast}) -- but for a shorter timescale of the order of some seconds. We attribute the latter to an overpressure induced by thermal transpiration and the continuous ejections to photophoretic forces as suggested by \citep{wurm2006}. Mass loss rates can be up to kg s$^{-1}$ for the continuous ejections and  $10^{-2}$ kg s$^{-1}$ due to ejections at the terminator for a 100 m body rotating once every 10 h (Fig.\ref{fig:planet_sketch}). The optical thick dust devils on Mars might also be supported by the two ejection mechanism, especially as the pressure is in the mbar regime and the surficial dust layers seem to be weakly-bound dust aggregates.

\section*{Acknowledgments}
This work is funded by the Deutsche Forschungsgemeinschaft. M.K. and J.K. are supported by the Scientific Grant Agency VEGA (Grants No. 2/0016/09), and by the Slovak Academy of Sciences (Grant No. 350/OMS/Fun/07; DAAD Project Nr. D0701266).




\section*{References}

\bibliographystyle{elsart-harv.bst}
\bibliography{kelling2011_icarus}



%
\end{document}